\documentclass[twocolumn,prl,superscriptaddress]{revtex4-2}
\usepackage[utf8]{inputenc}
\usepackage{times}
\usepackage{graphicx}
\usepackage[usenames,dvipsnames]{xcolor}  
\usepackage{amsmath}
\usepackage{amssymb}
\usepackage{bm}
\usepackage{hyperref}
\usepackage{siunitx}
\usepackage{listings}

\begin{document}

\title{Dislocation Glides in Granular Media}
\author{Fumiaki Nakai}
\email{fumiaki.nakai@ess.sci.osaka-u.ac.jp}
\affiliation{Department of Earth and Space Science, Osaka University, 1-1 Machikaneyama, Toyonaka 560-0043, Japan}
\author{Takashi Uneyama}
\affiliation{Department of Materials Physics, Graduate School of Engineering, Nagoya University, Furo-cho, Chikusa, Nagoya 464-8603, Japan}
\author{Yuto Sasaki}
\affiliation{Department of Earth and Space Science, Osaka University, 1-1 Machikaneyama, Toyonaka 560-0043, Japan}
\author{Kiwamu Yoshii}
\affiliation{Department of Physics, Graduate School of Science, Nagoya University, Furo-cho, Chikusa, Nagoya 464-8602, Japan}
\author{Hiroaki Katsuragi}
\affiliation{Department of Earth and Space Science, Osaka University, 1-1 Machikaneyama, Toyonaka 560-0043, Japan}

\begin{abstract}
Atomic crystals with dislocations deform plastically at low stresses via dislocation glide.
Whether dislocation glide occurs in macroscopic frictional granular media has remained unknown.
The discrete element method is employed to simulate the structural and mechanical responses of a granular crystal with an edge dislocation.
We find that dislocation glide occurs at low interparticle friction, resulting in significantly lower yield stresses than in dislocation-free crystals.
Yield stress varies linearly with interparticle friction, attributed to both Peierls stress and frictional effect.
\end{abstract}

\maketitle

{\it Introduction ---}
Microscopic crystals with line defects known as dislocations undergo plastic deformation at stress levels several orders of magnitude lower than those predicted by uniform deformation models~\cite{Peierls1940-ah, Nabarro1997-ry, Nabarro1997-sq, Joos1997-zn, Joos2001-au, Wang1996-kl, anderson2017theory}.
This low stress, known as the Peierls stress, results from successive local glides of dislocation sites within the crystal, a process analogous to inchworm movement~\cite{anderson2017theory, Bulatov2006-za}.
Dislocations are classified as edge dislocations when the dislocation line is perpendicular to the Burgers vector and as screw dislocations when these elements are parallel.
Both edge and screw dislocation glides can result in a wide range of plastic deformations~\cite{anderson2017theory, wagner1992molecular}.
Based on its physical origin, dislocation glide is not necessarily limited to microscopic systems. Nevertheless, in macroscopic systems consisting of particles several millimeters or larger in diameter, dislocation glides have been rarely observed, except in specific cases.

Bragg and Nye~\cite{bragg1947dynamical,bragg1949dynamical}, along with Nye and Lomer~\cite{lomer1949dynamical,lomer1952dynamical}, created a macroscopic two-dimensional crystal using numerous bubbles with uniform diameters of approximately \SI{0.5}{\milli\meter} floating on water.
Under large deformation, edge dislocations in this bubble crystal exhibit gliding behavior similar to that observed in microscopic crystals (screw dislocations are geometrically impossible in two-dimensional systems).
Subsequent studies have further confirmed dislocation glide in bubble systems~\cite{Vecchiolla2019-gv, Rosa1998-oc}.
Despite these observations in bubble systems, dislocation glide has not been reported in other macroscopic systems, even though the deformation of macroscopic granular crystals has been extensively studied~\cite{Karuriya2023-me, Panaitescu2012-pz, Otsuki2010-jk, Otsuki2023-vx, alam2003first, merkel2017enhanced}.
As such, the conditions necessary for the emergence of dislocation glide in macroscopic granular systems have remained unclear.
Given that bubbles floating on water can be considered a frictionless granular medium (interparticle friction $\mu_p=0$), a key question arises: Does dislocation glide emerge in frictional granular media ($\mu_p > 0$)?
If so, how does interparticle friction affect the rheological behaviors in granular crystals with a dislocation?

\begin{figure}
    \centering
    \includegraphics[width=0.47\textwidth]{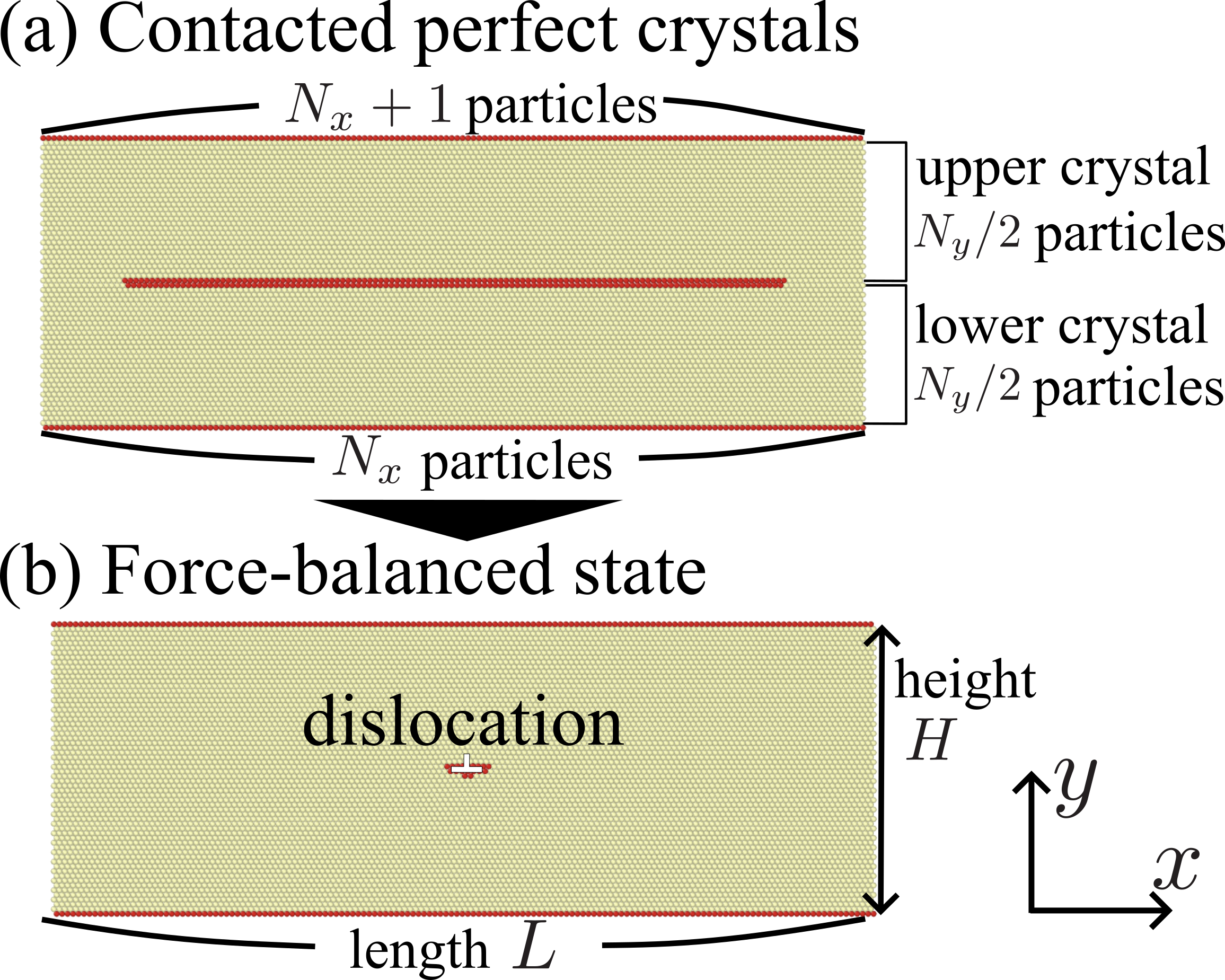}
    \caption{
    Setup of the system composed of frictional particles.
    (a) Initial arrangement of particles forming a two-dimensional hexagonal crystal structure. The upper crystal contains $(N_x+1)\times N_y/2$ particles, while the lower crystal contains $N_x\times N_y/2$ particles. $N_x$ and $N_y$ are set to 150 and 60, respectively.
    Particles with six contacting neighbors are colored light yellow; all others are colored red.
    Periodic boundary conditions are applied along the $x$-axis. The top and bottom layers consist of wall particles, while the rest are bulk particles confined between these walls.
    The system width is $L=N_x \alpha d$, where $\alpha$ is the lattice distance parameter, set to $0.95$.
    The system height $H$, defined as the distance between the centers of the top and bottom wall particles, is set to $H=\sqrt{3} (N_y+1)\alpha d/2$.
    (b) The system unloaded with fixed height $H$ and width $L$ under frictionless conditions ($\mu_p=0$), achieving a force-balanced state. This process creates a single dislocation at the center described using a symbol $\perp$ in the snapshot.
    Once the force-balanced state is achieved, interparticle friction $\mu_p$ is set to a target value, and shear is applied by moving the top wall along the $x$-axis while maintaining fixed $H$ and $L$.
    }
    \label{fig:initial}
\end{figure}

To address this issue, we numerically investigate the plastic deformation of a two-dimensional hexagonal crystal containing a single dislocation. The crystal is composed of frictional granular particles with size $d$ and mass density $\rho$, and we employ the conventional discrete element method (DEM)~\cite{Cundall1979-px, Mindlin2021-lh, Tsuji1992-xq, Poschel2003-lx, luding2008cohesive}.
We apply a constant shear rate $\dot{\gamma}$ and observe the structural changes and rheological responses of the granular crystal under various interparticle friction coefficients $\mu_p$.
For simplicity, we limit the study to slow deformations, where the dimensionless shear rate (also known as the inertia number) $\dot{\gamma}d/\sqrt{|\sigma_{yy}|/\rho}$ is less than $10^{-6}$. Here, $\sigma_{xy}$ and $\sigma_{yy}$ represent the shear and normal stresses, respectively.
Our results show that dislocation glide emerges for small interparticle friction coefficients ($\mu_p \lesssim 0.1$).
Conversely, at large $\mu_p$ values, dislocation glide does not occur as the crystal structure breaks down.
When dislocation glide is present, the yield stress $\sigma_{Y}$ can be significantly lower than that of a perfect crystal.
We find that $\sigma_{Y}$ is a linear function of $\mu_p$, which may be attributed to the combined effects of Peierls stress and interparticle friction.
The observation of dislocation glide in granular media opens new avenues for the rheological physics of granular materials, fostering interdisciplinary research that bridges the gap between microscopic crystals and macroscopic granular media.

{\it Simulation details ---}
We simulate dynamics of frictional granular particles under shear using the discrete element method (DEM)~\cite{Cundall1979-px,luding2008cohesive} implemented in LAMMPS (Large-scale Atomic/Molecular Massively Parallel Simulator), an open-source molecular dynamics program from Sandia National Laboratories~\cite{thompson2022lammps}.
As shown in Fig.~\ref{fig:initial}, the system consists of $(2N_x+1)N_y/2$ bulk particles with periodic boundary conditions along the $x$-axis. 
Along the $y$-axis, the bulk particles are bounded by parallel walls composed of identical particles.
The particle properties and material parameters such as diameter $d$, Young's modulus $E$, and Poisson's ratio $\nu$ are listed in Table~\ref{table:parameters}, which are set with reference to elastomer properties~\cite{callister2020materials,Robertson2007-sk}.
The inertia tensor $\bm{I}$ is determined assuming a uniform mass density distribution within each particle's volume.
We employ the conventional Hertz-Mindlin-Tsuji contact model~\cite{Mindlin2021-lh,Tsuji1992-xq} to calculate particle interactions, which are represented using Newton's equations of motion:
\begin{align}
    m \ddot{\bm{r}}_i &= \sum_{i\neq j} \bm{F}_{ij}=
    \sum_{i \neq j} (F_{n,ij}\bm{n}_{ij} + F_{\tau,ij}\bm{\tau}_{ij}) \Theta(d - |\bm{r}_{ij}|) ,\label{eq:motion_translational}\\
    \bm{I} \cdot \dot{\bm{\omega}}_i &=
    \sum_{i \neq j}
    (\bm{l}_{ij}\times \bm{\tau}_{ij})F_{\tau,ij}
    \Theta(d - |\bm{r}_{ij}|) , \label{eq:motion_rotational}
\end{align}
where $\bm{r}_{i}$ and $\bm{\omega}_i$ are the position and angular velocity of the $i$th particle, respectively.
$\bm{r}_{ij}$ is the relative position defined as $\bm{r}_{ij}=\bm{r}_i-\bm{r}_j$.
$F_{n,ij}$ and $F_{\tau,ij}$ are the normal and tangential forces between the $i$th and $j$th particles, respectively.
$\bm{l}_{ij}$ denotes the vector from the center of the $i$th particle to the contact point with the $j$th particle, while $\bm{n}_{ij}$ and $\bm{\tau}_{ij}$ represent the unit vectors of the normal and tangential components of the contact force between the $i$th and $j$th particles, respectively.
$\Theta(x)$ denotes the Heaviside step function of $x$.
$F_{n,ij}$ and $F_{\tau,ij}$ are given by
\begin{align}
    F_{n,ij} &= k_n\xi^{3/2}_{n,ij}
    + \eta v_{n,ij}, \label{eq:force_normal}\\
    F_{\tau,ij} &= 
    \min\left(\left|k_{\tau} \sqrt{\xi_{n,ij}}
    \bm{\xi}_{\tau,ij} + \eta \bm{v}_{\tau,ij}\right|, 
    \mu_p  F_{n,ij}\right),
    \label{eq:force_tangential}
\end{align}
where ${v}_{n,ij}=-\bm{v}_{ij}\cdot \bm{n}_{ij}$ and $\bm{v}_{\tau,ij}=\bm{v}_{ij}-\bm{v}_{ij}\cdot \bm{n}_{ij}\bm{n}_{ij} - \frac{d}{2}(\bm{\omega}_i+\bm{\omega}_j)\times \bm{n}_{ij}$, with the relative velocity $\bm{v}_{ij}=\dot{\bm{r}}_i-\dot{\bm{r}}_j$, denote the normal and tangential components of the relative velocity at the contact point between the $i$th and $j$th particles.
The spring constants are set as $k_n=\frac{E\sqrt{d}}{3(1-\nu^2)}$ and $k_{\tau}=\frac{E\sqrt{d}}{(1+\nu)(2-\nu)}$.
Following conventions~\cite{Tsuji1992-xq,Silbert2001-zj,Marshall2009-yp}, the damping constants for normal and tangential forces are set to the same value: $\eta=\kappa \sqrt{m k_n\sqrt{\xi_{n,ij}}/2}$, where $\kappa=1.2728-4.2783e+11.087e^2-22.348e^3+27.467e^4-18.022e^5+4.8218e^6$. This formulation maintains a constant restitution coefficient $e$.
$\bm{\xi}_{\tau,ij}$ denotes the accumulated tangential displacement vector (see details in \cite{luding2008cohesive}).
$\xi_{n,ij}$ is the overlap length between the $i$th and $j$th particles, defined as $d-|\bm{r}_{ij}|$.
The $\min$ function in Eq.~\eqref{eq:force_tangential} implements Coulomb's law of friction, where $\mu_p$ represents the interparticle friction coefficient. For simplicity, the static and kinetic friction coefficients are assumed to be equal.
For the calculation of the equation of motion based on Eqs.~\eqref{eq:motion_translational} and \eqref{eq:motion_rotational}, the time step size is set to $\SI{8.9e-6}{\second}$.

\begin{table}[htbp]
\caption{Parameters for the DEM simulation with the Hertz/Mindlin/Tsuji contact model. Material constants are based on elastomer properties~\cite{callister2020materials, Robertson2007-sk}.\label{table:parameters}}
\begin{tabular}{c|c|c}
\hline
Parameter & Symbol & Value \\ \hline
Mass density of particle & $\rho$ & $\SI{1000}{\kg\per\m^3}$\\
Diameter of particle & $d$ & \SI{1}{\mm}\\
Young's modulus of particle & $E$ & \SI{1}{\MPa}\\
Poisson's ratio of particle & $\nu$ & 0.45\\
Restitution coefficient of particle & $e$ & 0.6\\
Lattice distance parameter & $\alpha$ & 0.95\\
Number of particles & $N$ & $(2N_x+1) N_y/2$\\
Side length of box & $L$ & $N_x\alpha d$\\
Height of box & $H$ & $\sqrt{3}(N_y+1)\alpha d/2$\\
Shear rate & $\dot{\gamma}$ & \SI{e-3}{\per\second}\\
Interparticle friction coefficient & $\mu_p$ & $0\le \mu_p\le 1$\\
\hline
\end{tabular}
\end{table}

To generate a dislocation in the crystal, we construct the initial state based on a conventional method for atomic dislocations~\cite{anderson2017theory, osetsky2003atomic}.
As presented in Fig.~\ref{fig:initial}, $N_x$ wall particles and $N_x\times \frac{N_y}{2}$ bulk particles in the lower half are positioned to construct a hexagonal crystal with the lattice vectors $(\alpha d, 0)$ and $\left( \frac{\alpha d}{2}, \frac{\sqrt{3}\alpha d}{2}\right)$, where $\alpha = 0.95$.
For the $N_x+1$ wall particles and $(N_x+1)\times \frac{N_y}{2}$ bulk particles in the upper half, the lattice vectors are $\left( \frac{N_x}{N_x+1}\alpha d, 0\right)$ and $\left(\frac{N_x}{N_x+1}\frac{\alpha d}{2}, \frac{\sqrt{3}\alpha d}{2}\right)$.
This preparation causes a misalignment of one lattice unit across the interface between the upper and lower side crystals, as shown in Fig.~\ref{fig:initial}(a).
We create a force-balanced state based on Eqs.~\eqref{eq:motion_translational} and $\eqref{eq:motion_rotational}$ under the frictionless condition ($\mu_p=0$), while fixing the motion of the wall particles along the $y$-axis and keeping the distances between neighboring wall particles constant.
This procedure generates a dislocation at the interface between the upper and lower side particles, represented by a symbol $\perp$ in Fig.~\ref{fig:initial}(b).
From this force-balanced state, we set a target $\mu_p$ and apply a constant shear rate $\dot{\gamma}=\SI{e-3}{\per\s}$ (inertia number becomes $\dot{\gamma}d/\sqrt{|\sigma_{yy}|/\rho}<\SI{e-6}{}$ in the current setup) by moving the top wall with a constant velocity $\dot{\gamma}H$ along the $x$-axis.
We set the initial time $t=0$ at the moment when the shear is applied and analyze the subsequent data for various values of $\mu_p\in [0,1]$.

\begin{figure}
    \centering
    \includegraphics[width=0.47\textwidth]{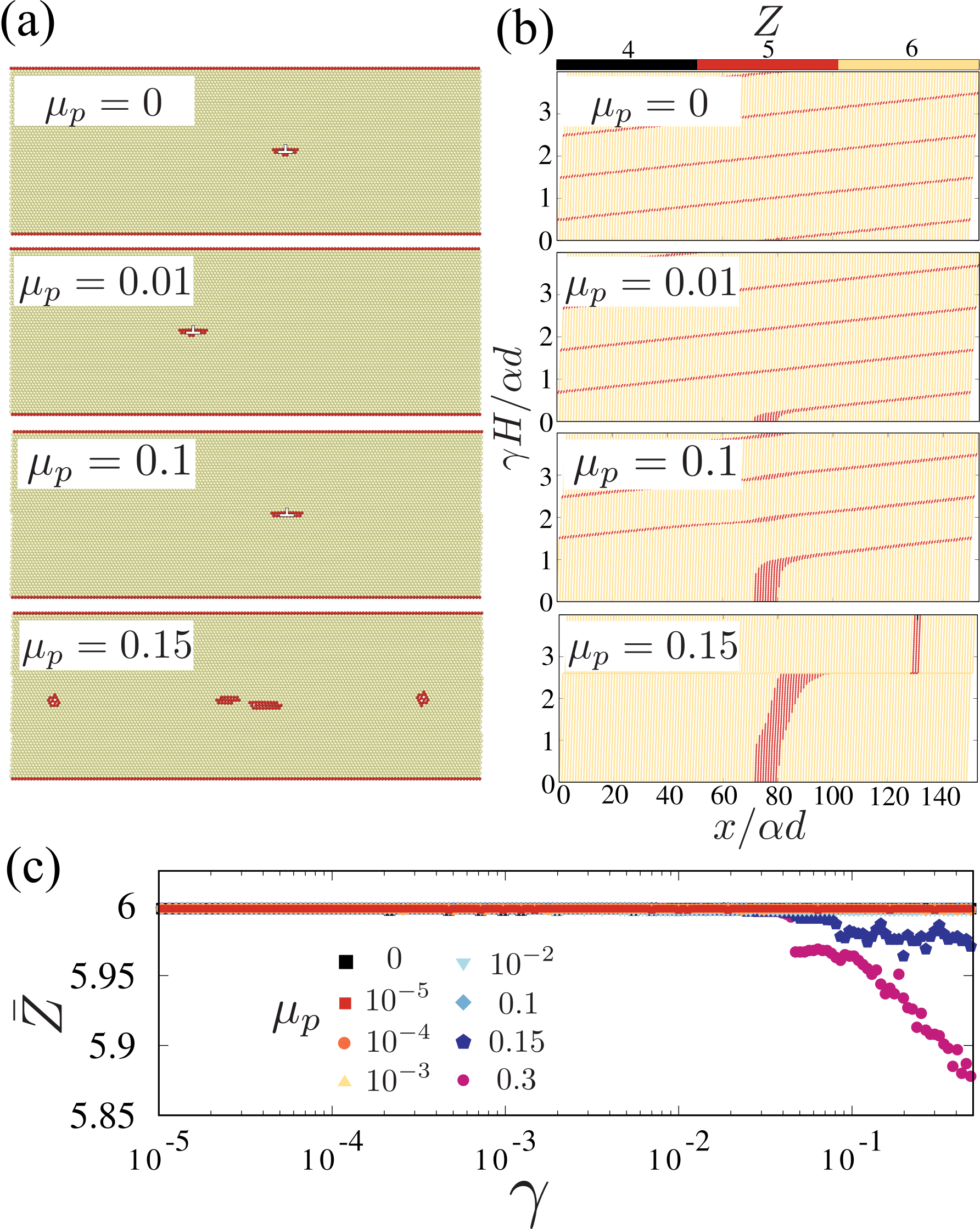}
    \caption{
    (a) Snapshots taken after applying a small strain of $\gamma H/\alpha d=4$ ($\gamma\simeq 0.077$), shown for various values of interparticle friction $\mu_p$.
    Yellow particles are in contact with six other particles, while red particles are in contact with fewer than six.
    For $\mu_p \lesssim 0.1$, crystal order is preserved due to dislocation glide.
    At higher $\mu_p$ values, gliding ceases, leading to partial destruction of crystal order.
    (b) Trajectories of $N_x+1$ particles in the upper crystal, initially located at the interface. Horizontal and vertical axes represent the scaled particle positions $x/\alpha d$ and scaled strain $\gamma H/\alpha d$. 
    The color indicates the number of particles in contact, denoted by $Z$.
    For $\mu_p \lesssim 0.1$, the red region moves straight with a displacement $\lambda = \gamma HL/\alpha d$ relative to strain $\gamma$.
    This behavior corresponds to dislocation glide, which does not occur at $\mu_p \gtrsim 0.15$ due to crystal breakage.
    (c) Average number of particles in contact per particle, denoted by $\bar{Z}$, as a function of $\gamma$.
    $\bar{Z}$ equals $6$ for a hexagonal structure and deviates from $6$ otherwise.
    When $\mu_p \lesssim 0.1$, $\bar{Z}$ remains 6, indicating that the hexagonal structure is stable under shear.
    For high interparticle friction ($\mu_p \gtrsim 0.15$), $\bar{Z}$ decreases with increasing $\gamma$, indicating the breakage of hexagonal order.}
    \label{fig:structure}
\end{figure}

{\it Results ---}
The structural response to shear qualitatively depends on $\mu_p$ (see Fig.~\ref{fig:structure}(a)).
In the frictionless case ($\mu_p=0$), as expected, the dislocation site glides successively without disrupting the crystal order.
This gliding behavior emerges even for weak interparticle friction $\mu_p \lesssim 0.1$.
The observed dislocation glide appears similar to that on the atomic scale~\cite{lunev2017glide}.
In contrast, when $\mu_p \gtrsim 0.15$, the hexagonal structure partially breaks, and the dislocation glide does not occur.
The trajectories of the upper-side particles initially located at the interface exhibit different types of motion with respect to $\mu_p$, as shown in Fig.~\ref{fig:structure}(b).
The color represents the number of contacting particles $Z$, which equals $6$ when the local structure around a particle is hexagonal; otherwise, $Z \neq 6$.
For $\mu_p \lesssim 0.1$, a continuous red line is observed, corresponding to the dislocation glide.
The dislocation site moves over the system size with the period $\gamma H/\alpha d=1$, which does not depend on $\mu_p$.
This period agrees with Orowan's theory~\cite{anderson2017theory}, which is a simple geometrical consequence where the dislocation site moves by the system length $L$ relative to the wall displacement of a single lattice distance $\alpha d$. The displacement of the dislocation glide $\lambda$ relative to $\gamma$ is
\begin{equation}
    \lambda = \frac{\gamma H L}{\alpha d}.
    \label{eq:orowan}
\end{equation}
where $\lambda=L$ leads to $\gamma H/\alpha d=1$.
This consistency between our results and Orowan's theory confirms that the observed motion is indeed the dislocation glide.
For $\mu_p \geq 0.15$, the continuous red line does not appear as dislocation glide is impeded by crystal destruction (see Fig.~\ref{fig:structure}(a)).
To quantify the bulk structure, the average number of contacting particles per particle, $\bar{Z}$, is computed as a function of strain $\gamma$ (see Fig.~\ref{fig:structure}(c)).
When $\mu_p \lesssim 0.1$, $\bar{Z}$ remains close to $6$ even for large $\gamma$, indicating that the hexagonal order is stable under shear due to dislocation glide, as shown in Fig.~\ref{fig:structure}(a).
Conversely, when $\mu_p \gtrsim 0.15$, $\bar{Z}$ deviates from $6$ with increasing $\gamma$, indicating the breakage of the hexagonal structure.
These results demonstrate that the current setup exhibits dislocation glide when $\mu_p \lesssim 0.1$.
This work presents the first observation of dislocation glide in frictional granular media.

\begin{figure}
    \centering
    \includegraphics[width=0.47\textwidth]{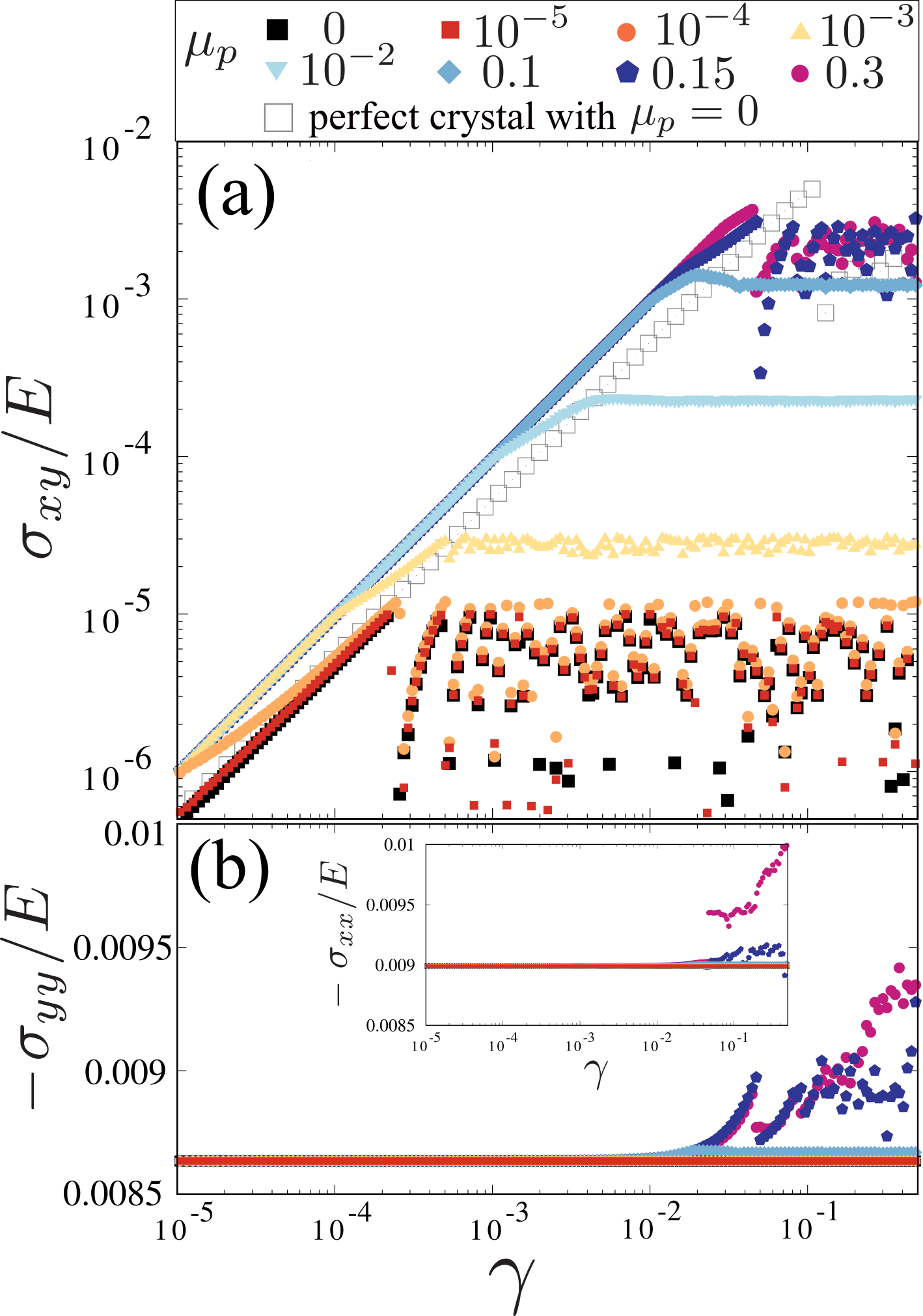}
    \caption{
    Rheological responses for various interparticle friction coefficients $\mu_p$.
    (a) The normalized shear stress $\sigma_{xy}/E$ as a function of shear strain $\gamma$. Simulation data for the frictionless perfect crystal is also shown for comparison.
    For small $\gamma$, linear responses are observed for all $\mu_p$ values.
    Yielding behavior, defined as the first peak of $\sigma_{xy}$, occurs at a specific $\gamma$.
    The yield stress of the crystal with the dislocation decreases as $\mu_p$ decreases and can be significantly lower than that of the perfect crystal.
    (b) The normalized normal stress $-\sigma_{yy}/E$ as a function of $\gamma$.
    The increase in $-\sigma_{yy}$ with $\gamma$, corresponding to dilatancy, is not observed for small interparticle friction ($\mu_p \lesssim 0.1$) due to dislocation glide.
    The inset shows $-\sigma_{xx}/E$ vs $\gamma$, which also does not exhibit dilatant behavior for $\mu_p \lesssim 0.1$.
    }
    \label{fig:stress}
\end{figure}

The rheological behavior is expected to change qualitatively in response to structural changes induced by variations in $\mu_p$.
We compute the stress tensor of the bulk, which is defined as follows~\cite{Allen2017-ja, J_Evans2007-di}:
\begin{equation}
    \bm{\sigma} =
    -\frac{1}{dLH}\sum_{i=1}^{N}\left[
    m\dot{\bm{r}}_i \dot{\bm{r}}_i
    +\frac{1}{2}\sum_{j\ne i}^{N}\bm{r}_{ij} \bm{F}_{ij}\right],
    \label{eq:stress}
\end{equation}
where $m$ denotes the mass of the particle, and $\bm{F}_{ij}$ has been defined in Eq.~\eqref{eq:motion_translational}.
In the current setup with strain rate $\dot{\gamma}=\SI{e-3}{\per\s}$, the contribution of the kinetic term in Eq.~\eqref{eq:stress} is negligibly small.
Fig.~\ref{fig:stress}(a) shows the logarithmic plot of shear stress $\sigma_{xy}$ normalized by $E$ as a function of strain $\gamma$. For reference, $\sigma_{xy}$ of the frictionless perfect crystal ($N=N_x\times N_y$ with $\alpha=0.95$) is also presented.
For all values of $\mu_p$, linear responses ($\sigma_{xy}\propto \gamma$) are observed at small $\gamma$, and yielding behaviors occurs at specific $\gamma$.
For frictionless case ($\mu_p=0$) with the dislocation, $\sigma_{xy}$ exhibits extremely low yield stress, which may be characterized as the first peak in $\sigma_{xy}$, compared to that of the perfect crystal. This stress, known as the Peierls stress~\cite{anderson2017theory, Peierls1940-ah, Nabarro1997-ry, Nabarro1997-sq, Joos1997-zn, Joos2001-au}, originates from dislocation glide.
For finite but small interparticle friction ($\mu_p \lesssim 0.1$), where dislocation glide occurs (see Fig.~\ref{fig:structure}(a)), both yield stress and strain increase monotonically with $\mu_p$.

Granular media often exhibit dilatancy, a volume expansion under shear~\cite{Andreotti2013-yo}.
In our constant volume setup, dilatancy can be characterized by an increase in $-\sigma_{yy}$ with respect to $\gamma$, as shown for $\mu_p \gtrsim 0.15$ in Fig.~\ref{fig:stress}(b).
A similar trend for $\sigma_{xx}$ is shown in the inset.
However, for small interparticle friction ($\mu_p \lesssim 0.1$), no dilatancy is observed, with $\sigma_{yy}$ and $\sigma_{xx}$ remaining constant.
This absence of dilatancy can be intuitively understood, as the system deformation is relaxed by dislocation glide before the onset of dilatancy.

\begin{figure}
    \centering
    \includegraphics[width=0.47\textwidth]{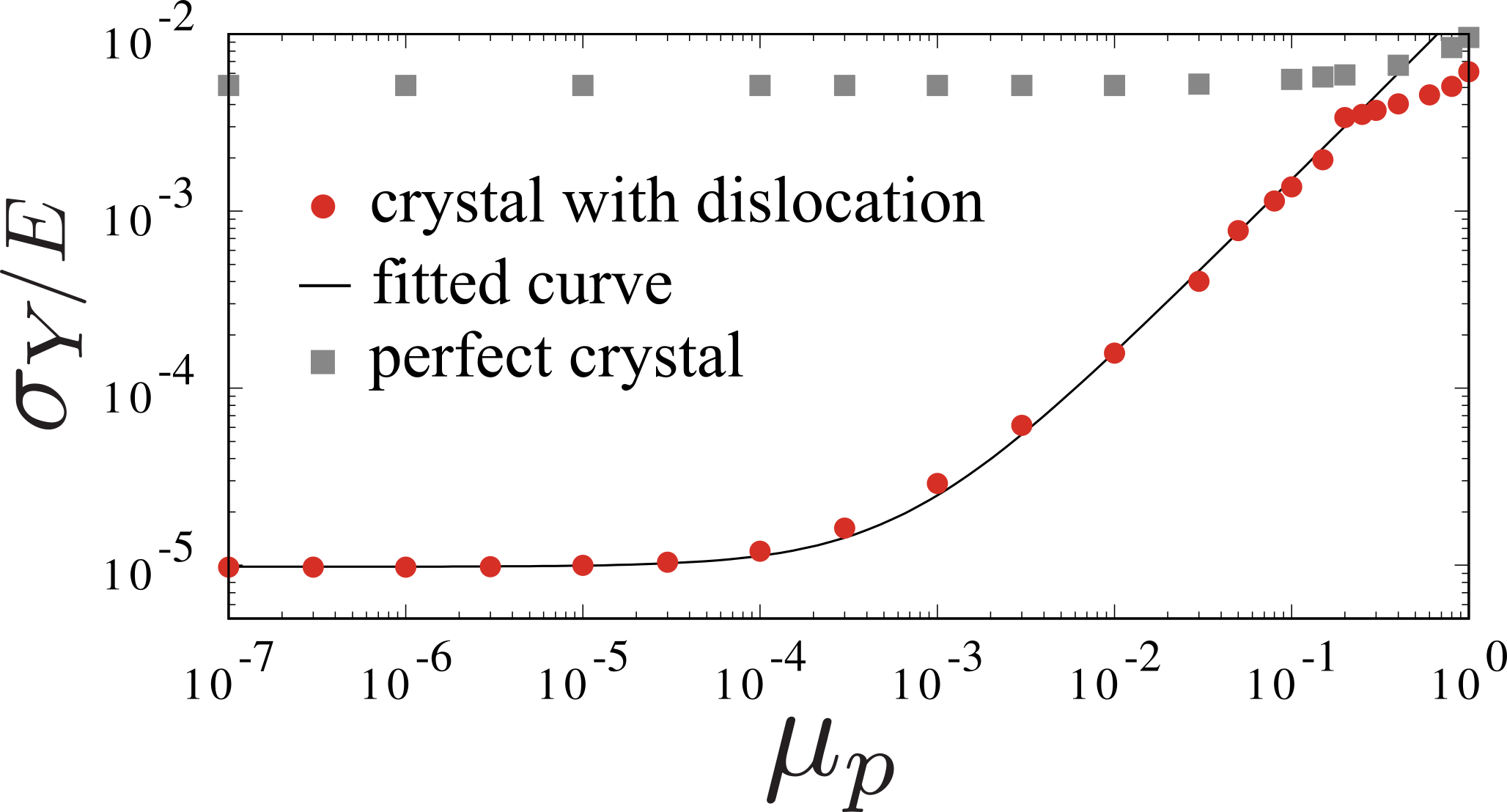}
    \caption{
    Yield stress $\sigma_{Y}$ of the crystal with dislocation, defined as the first peak top value in $\sigma_{xy}$ vs $\gamma$, as a function of $\mu_p$.
    Data for the perfect crystal with finite $\mu_p$ are also shown for comparison.
    Due to dislocation glide, $\sigma_Y$ of the crystal with the dislocation can be significantly lower than that of the perfect crystal.
    For $\mu_p \lesssim 0.1$, $\sigma_Y$ of the crystal with a dislocation can be fitted by the linear function $\sigma_Y/E = \SI{9.8e-6}{} + \SI{1.5e-2}{}\mu_p$. This relation arises from contributions of both the Peierls stress and interparticle friction.
    }
    \label{fig:yield}
\end{figure}

Fig.~\ref{fig:yield} shows the yield stress $\sigma_Y$ of the crystal with the single dislocation as a function of $\mu_p$, where $\sigma_Y$ is defined as the first peak in the $\sigma_{xy}$ vs $\gamma$ plot.
For comparison, we also include $\sigma_Y$ of the perfect crystal with finite $\mu_p$, containing $N_x \times N_y$ particles with the lattice distance parameter $\alpha=0.95$.
For $\mu_p \lesssim 0.1$, $\sigma_Y$ of the crystal with the dislocation exhibits a linear relation with $\mu_p$, which can be fitted by $\sigma_Y/E=\SI{9.8e-6}{}+\SI{1.5e-2}{}\mu_p$.
For $\mu_p \gtrsim 0.1$, the data deviates from this linear relation as the crystal order breaks down and dislocation glide ceases, as shown in Fig.~\ref{fig:structure}(a)-(c).
Notably, due to dislocation glide, $\sigma_Y$ of the crystal with the dislocation can be significantly lower than that of the perfect crystal.

The linear relation between $\sigma_Y$ and $\mu_p$ can be explained by considering both the Peierls stress and interparticle friction.
In the frictionless case, the dislocation moves by overcoming a potential barrier, with the required stress being the Peierls stress $\sigma_{\rm{Peierls}}$ \cite{Peierls1940-ah, Nabarro1997-ry, Nabarro1997-sq, Joos1997-zn, Joos2001-au, Wang1996-kl, anderson2017theory}. (Note that despite numerous attempts~\cite{Peierls1940-ah,Nabarro1997-ry,anderson2017theory,Joos1997-zn}, precise theoretical estimation of the Peierls stress remains challenging and is beyond the scope of this work.)
In our frictional case, we must also consider the effect of interparticle friction, which is absent in microscopic crystals.
Dislocation glide involves local rearrangement between contacting particles at the dislocation site. In the presence of friction, the particles must also overcome Coulomb's friction criterion to change their configuration.
The normal stress in the system is approximately $|\sigma_{xx}|/E\simeq |\sigma_{yy}|/E \simeq \SI{e-2}{}$ (see Fig.~\ref{fig:stress}(b)).
The local stress required for rearrangement between contacting particles at the dislocation site can be approximately estimated as the product of $\mu_p$ and the normal stress: $\mu_p|\sigma_{yy}|\simeq \SI{e-2}{}\mu_p E$.
Combining the contributions from the Peierls stress and interparticle friction provides the linear relation $\sigma_Y \simeq \sigma_{\rm{Peierls}} + \mu_p|\sigma_{yy}|$, which is consistent with the observations in Fig.~\ref{fig:yield}.

{\it Summary ---}
This study numerically investigates the structural and rheological responses of a granular crystal containing a single dislocation under slow shear conditions (inertia number $\dot{\gamma}d/\sqrt{|\sigma_{yy}|/\rho} < 10^{-6}$).
Our findings show that for low interparticle friction ($\mu_p \lesssim 0.1$), dislocation glide occurs while maintaining crystal symmetry. In contrast, for high $\mu_p$, the crystal structure breaks down, impeding dislocation glide.
For $\mu_p \lesssim 0.1$, the yield stress $\sigma_Y$ can exhibit significantly lower values compared to those of the perfect crystal, with the observed linear relation between $\sigma_Y$ and $\mu_p$ attributed to both the Peierls stress and interparticle friction.
The experimental realization of the setup employed in this study may be feasible, albeit requiring meticulous preparation.
Future work should focus on analyzing the effects of particle size distribution, shape anisotropy, and spatial dimensions to explore the experimental limitations.
Extending this work through investigations into screw dislocations, multi-dislocation systems, and dislocation creation/annihilation, which has been deeply investigated in atomic crystals, will be promising research directions.
This study bridges the physics of granular media and atomic crystals, offering interdisciplinary insights.

\section*{Acknowledgement}
F.N. was supported by Grant-in-Aid for JSPS (Japan
Society for the Promotion of Science) Fellows (Grant No. JP24KJ0156).
The computation in this work has been done using the facilities of the Supercomputer Center, the Institute for Solid State Physics, the University of Tokyo.

\bibliographystyle{apsrev4-2}
\bibliography{ref}

\end{document}